\documentclass[twocolumn,secnumarabic,amssymb, nobibnotes, aps, prd]{revtex4-2}

\usepackage{natbib}
\usepackage{amsmath}
\usepackage{graphicx}
\usepackage{xcolor}
\usepackage{dsfont}
\usepackage[normalem]{ulem}

\newcommand{\bRcal}{\boldsymbol{\mathcal{R}}}

\begin{document}
\title{\emph{Ab initio} study of  metastable occupation of tetrahedral sites in palladium hydrides and its impact on superconductivity}

\author{Antonella Meninno$^{1,2}$}\email[]{ameninno001@ikasle.ehu.es}\author{ Ion Errea$^{1,2,3}$}

\affiliation {$^1$Centro de F\'isica de Materiales (CSIC-UPV/EHU), Manuel de Lardizabal Pasealekua 5, 20018 Donostia/San Sebasti\'an, Spain}

\affiliation {$^2$Fisika Aplikatua Saila, Gipuzkoako Ingeniaritza Eskola, University of the Basque Country (UPV/EHU), Europa Plaza 1, 20018 Donostia/San Sebasti\'an, Spain} 

\affiliation {$^3$Donostia International Physics Center (DIPC), Manuel de Lardizabal Pasealekua 4, 20018 Donostia/San Sebasti\'an, Spain}

\begin{abstract}
A recent experimental work on palladium hydrides suggested that metastable structures with hydrogen atoms occupying tetrahedral sites could lead to superconductivity above 50 K, a huge increase compared to the 10 K critical temperature of the stable structure with all hydrogen atoms occupying octahedral sites. By generating many structures with hydrogen atoms randomly occupying the octahedral and tetrahedral sites of the face-centered cubic lattice and calculating their energy at different theoretical levels from first principles, we determine that metastable structures with partial or full occupation of tetrahedral sites are possible, even when the ionic quantum zero-point energy and anharmonicity are included in the calculations. Anharmonicity is crucial in palladium hydrides when hydrogen atoms occupy octahedral sites and, in fact, makes the structure with full octahedral occupation the ground state. Despite the metastable existence of structures with full or partial tetrahedral sites occupation, the superconducting critical temperature is reduced with the number of tetrahedral sites occupied. Our calculations discard that the occupation of tetrahedral sites can increase the critical temperature in palladium hydrides.
\end{abstract}

\maketitle

\section{Introduction}

\label{intro}

The discovery of superconductivity above 200 K at megabar pressures in hydrogen-based compounds, beating all the records previously set by the cuprates, is one of the greatest discoveries in physics of the last years \cite{drozdov2015conventional,somayazulu2019evidence,drozdov2019superconductivity,snider2021synthesis,kong2021superconductivity,troyan2021anomalous}. The question whether stable or metastable hydrides with a high superconducting critical temperature ($T_c$) exist also at ambient pressure has to be answered now. There are hydrogen-based superconductors at ambient pressure such as PdH \cite{skoskiewicz1972superconductivity,stritzker257,schirber1974concentration,schirber1984superconductivity} and Th$_4$H$_{15}$ \cite{satterthwaite1970superconductivity}, but all of them have low $T_c$ values, not surpassing $\sim$ 10 K. Interestingly, Syed et al., in a work that remains unconfirmed, have claimed that metastable phases of PdH grown using a fast-cooling technique can yield much higher superconducting critical temperatures, even reaching values above 50 K \cite{syed2016superconductivity}. If confirmed, that would mean high-$T_c$ hydrides can exist at ambient pressure.

Palladium hydrides have been largely studied since their discovery \cite{graham1869relation}, for instance, as hydrogen storage materials.  In fact, palladium has a large capacity  to absorb considerable quantities of hydrogen \cite{alefeld1978hydrogen,fukai2006metal,lewis1967palladium}. It is possible to synthesize compounds with different values of the H/Pd ratio, labeled as $x$, with proportions that reach as maximum the stoichiometric condition ($x=1$). PdH$_x$ becomes superconducting with $x>0.80$, and the critical temperature rises as $x$ is increased, reaching a value of 8-9 K in the stoichiometric limit \cite{skoskiewicz1972superconductivity,stritzker257,schirber1974concentration,schirber1984superconductivity}. Interestingly, the superconducting critical temperature of PdH increases the heavier the mass of the hydrogen \cite{schirber1974concentration,schirber1984superconductivity}, a strongly anomalous isotope effect that was explained due to the presence of strong anharmonic hydrogen lattice vibrations \cite{errea2013first}.

The general consensus is that the ground state structure of stoichiometric PdH is formed by hydrogen atoms occupying the octahedral interstitial sites of the face centered cubic (fcc) palladium lattice. This statement is supported by early neutron diffraction experiments both on the hydride and the deuteride \cite{WORSHAM1957neutron}, as well as the agreement between the theoretical \emph{ab initio} anharmonic phonon calculations \cite{errea2013first} and the experimental phonon spectra obtained with inelastic neutron scattering and Raman experiments \cite{rowe1974lattice,rowe1986isotope,Kolesnikov1991neutron,Sherman1977raman,ross1998strong,Chowdhury1973neutron}. The fact that the superconducting properties, including the inverse isotope effect, are explained by \emph{ab initio} theoretical calculations assuming the octahedral occupation further sustains this idea \cite{errea2013first}. However, more recent powder neutron diffraction experiments have observed  partial occupation of the interstitial tetrahedral sites in the deuteride compound \cite{Pitt2003Tetrahedral,McLennan2008deuterium}. First-principles calculations within density-functional theory (DFT) \cite{elsasser1991vibrational} show that both octahedral and tetrahedral configurations are local energy minima, where the octahedral configuration is slightly favored with respect to the tetrahedral one. Similar DFT calculations determine, on the contrary, that tetrahedral site occupation is preferred energetically at high hydrogen concentration \cite{caputo2003h}.
 
The question of where the hydrogen atoms sit in  PdH is of particular importance because the main hypothesis to explain the experiments in Ref. \cite{syed2016superconductivity} is that the fast cooling technique yielded metastable PdH and PdD compounds with partial or full tetrahedral site occupation and high superconducting critical temperatures. The energetic comparison between different possible sites has not been performed fully including the lattice zero-point energy and anharmonicity, which is  crucial in this system \cite{errea2013first}, beyond simplified models \cite{elsasser1991vibrational}. In order to clarify the possibility of having stable or metastable occupation of tetrahedral sites in PdH and its impact on superconductivity, here we present \emph{ab initio} structural relaxations of several PdH structures with mixed octahedral and tetrahedral site occupations including the quantum lattice zero-point energy within the stochastic self-consistent harmonic approximation (SSCHA) \cite{errea2014anharmonic,bianco2017second,monacelli2018pressure,Monacelli2021thestochastic}. Our results suggest that metastable full or partial tetrahedral site occupation of interstitial sites is possible in PdH, but that it does not enhance the superconducting critical temperature.   

The manuscript is organized as follows. In Sec. \ref{methods} we describe the details of the \emph{ab initio} calculations, in Sec. \ref{sec:results} we present the results of our work, and in Sec. \ref{sec:conclusions} we summarize the main conclusions.

\section{Methods and computational details}

\label{methods}

In the present analysis we work always with the stoichiometric ratio $x=1$. In order to study the energies of configurations with tetrahedral site occupation, we have constructed a $2\times2\times2$ supercell starting from the primitive fcc lattice, i.e. with 8 Pd atoms, and randomly place 8 hydrogen atoms between the octahedral and tetrahedral sites. We classify the structures by the number of occupied octahedral sites in the beginning. In any configuration, if  $n$ hydrogen atoms occupy octahedral sites (with $n$ from 0 to 8), then $8-n$ hydrogen atoms occupy tetrahedral sites. Note that per Pd atom in the fcc lattice there is one octahedral site and 2 tetrahedral sites. The structure with full octahedral occupation, $n=8$, has a space group symmetry $Fm\bar{3}m$ and all positions are fixed by symmetry. Configurations with partial or full tetrahedral occupation have a lower symmetry and, in most of the cases, atoms are no longer fixed to a specific lattice site and can thus relax. We have performed these relaxations both in the classical case, assuming that ions are classical objects and therefore adopt the positions given by the local minima of the Born-Oppenheimer energy surface $V(\mathbf{R})$, and in the quantum case within the SSCHA. 

The SSCHA is a variational method that minimizes the free energy 
\begin{equation}
    \mathcal{F}[\boldsymbol{\mathcal{R}},\boldsymbol{\Phi}] = \langle K + V(\mathbf{R})
    \rangle_{\tilde{\rho}_{\boldsymbol{\mathcal{R}},\boldsymbol{\Phi}}}-TS_{\mathrm{ion}}[\tilde{\rho}_{\boldsymbol{\mathcal{R}},\boldsymbol{\Phi}}]
    \label{eq:sscha_energy}
\end{equation}
of the system as a function of the \emph{centroid} positions $\boldsymbol{\mathcal{R}}$ and auxiliary force constants $\boldsymbol{\Phi}$ that parametrize the ionic density matrix $\tilde{\rho}_{\boldsymbol{\mathcal{R}},\boldsymbol{\Phi}}$ \cite{errea2014anharmonic,bianco2017second,monacelli2018pressure,Monacelli2021thestochastic}. In Eq. \eqref{eq:sscha_energy} $K$ is the ionic kinetic energy operator, $S_{\mathrm{ion}}[\tilde{\rho}_{\boldsymbol{\mathcal{R}},\boldsymbol{\Phi}}]$ the ionic entropy associated to $\tilde{\rho}_{\boldsymbol{\mathcal{R}},\boldsymbol{\Phi}}$, and $\langle \rangle_{\tilde{\rho}_{\boldsymbol{\mathcal{R}},\boldsymbol{\Phi}}}$ denotes the quantum statistical average taken with $\tilde{\rho}_{\boldsymbol{\mathcal{R}},\boldsymbol{\Phi}}$. The SSCHA assumes that the probability distribution function defined by $\tilde{\rho}_{\boldsymbol{\mathcal{R}},\boldsymbol{\Phi}}$ is a Gaussian centered at the $\boldsymbol{\mathcal{R}}$ positions and has a width related to $\boldsymbol{\Phi}$. At the minimum of $\mathcal{F}[\boldsymbol{\mathcal{R}},\boldsymbol{\Phi}]$, the obtained centroid positions $\boldsymbol{\mathcal{R}}$ determine the renormalized average positions of the ions including ionic quantum effects and anharmonicity at a non-perturbative level.

Classical structural relaxations on the Born-Oppenheimer energy surface were performed with DFT making use of the plane-waves {\sc Quantum ESPRESSO} package \cite{giannozzi2009quantum,giannozzi2017advanced}. The exchange-correlation functional was approximated with the Perdew-Burke-Ernzerhof (PBE) parametrization \cite{perdew1996generalized}. Brillouin zone integrals in the DFT self-consistent calculations were performed with a 12$\times$12$\times$12 $\mathbf{k}$-point grid for the $2\times2\times2$ supercell containing 16 atoms. We used a kinetic energy cutoff for the wave functions of 55 Ry and of 550 Ry for charge density. Projector-augmented wave pseudopotentials from the {\sc Quantum ESPRESSO} library were used \cite{DALCORSO2014337}, with 10 electrons in the valence for Pd.

\begin{figure}
	\includegraphics[width=\linewidth]{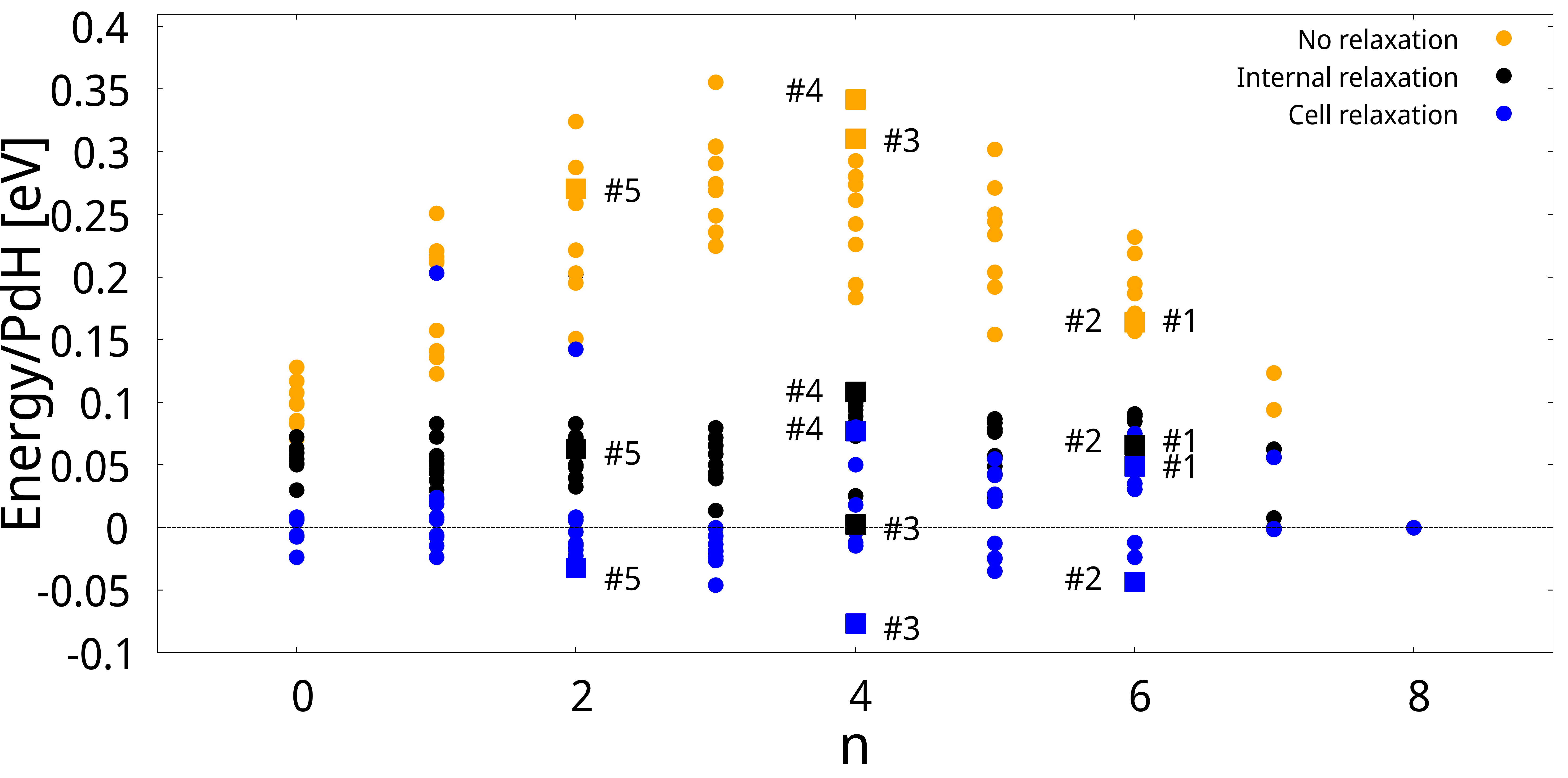}
	\includegraphics[width=\linewidth]{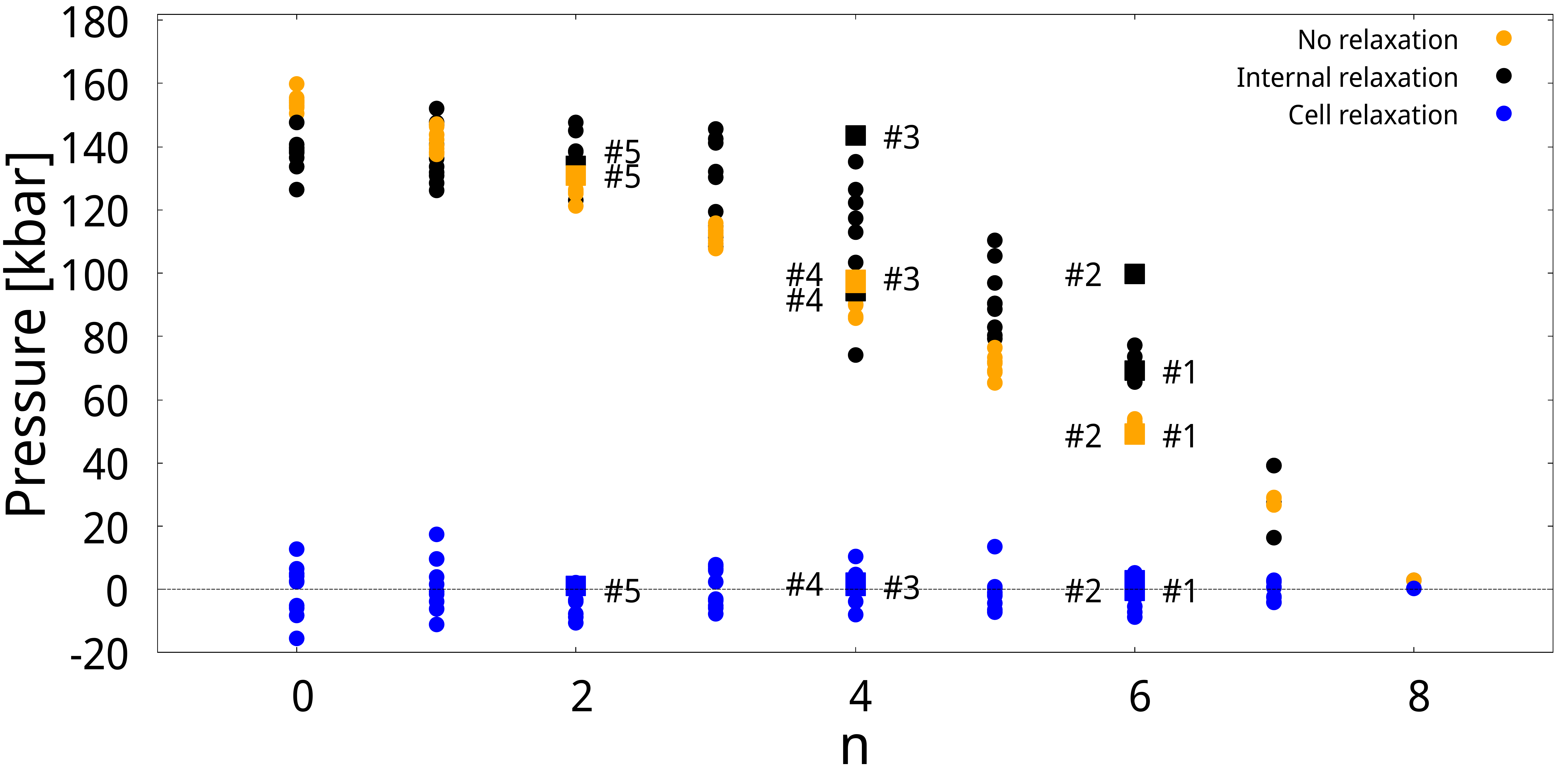}
	\caption{Born-Oppenheimer energies (top panel) and pressures (bottom panel) for structures  with hydrogen atoms at random interstitial sites. The structures are classified according to the number of occupied octahedral sites at the beginning of the calculation, $n$. The data in orange refers to the initial non-relaxed structures, while in black we show the data for the internal relaxed structures (internal relaxation) keeping the unit cell unchanged. The data in blue corresponds to the classical relaxation in which the unit cell is allowed to change (cell ralaxation). In each class of data, we have marked with numbers the energies and pressures of the structures that we study with the SSCHA method \emph{a posteriori}, which are marked with a square and not a dot. The energies are measured from the energy of the structure with full octahedral occupation, $n=8$.}
	\label{energies}
\end{figure}

SSCHA quantum anharmonic structural relaxations were performed on the same $2\times2\times2$ supercell with 16 atoms. The SSCHA minimization requires the calculation of atomic forces on different random configurations generated according to $\tilde{\rho}_{\boldsymbol{\mathcal{R}},\boldsymbol{\Phi}}$. These forces were calculated \emph{ab initio} with DFT, with the same parameters used for the classical relaxations. Due to the reduced symmetry of many of the structures analyzed, a large number of random configurations were needed to converge the SSCHA minimization, in the order of 10000 per compound for the less symmetric cases.

We have calculated the phonon spectra of the structures relaxed at the quantum level with the SSCHA both at the harmonic level and the anharmonic level. The electron-phonon interaction is also estimated at the harmonic and at the SSCHA levels for these structures. The harmonic calculations have been performed within density-functional perturbation theory (DFPT) \cite{baroni2001phonons} in a $2\times2\times2$ $\mathbf{q}$-point grid of the 16 atoms supercell. The harmonic Eliashberg function $\alpha^2F(\omega)$ \cite{giustino2017electron-phonon} has been obtained with electron-phonon matrix elements calculated in the $2\times2\times2$ $\mathbf{q}$-point grid, a $30\times 30 \times 30$ $\mathbf{k}$-point grid for the Brillouin zone integrals on the electronic states, and a 0.002 Ry Gaussian smearing for the double Dirac delta on the Fermi surface. In order to obtain the anharmonic $\alpha^2F(\omega)$ in the same $2\times2\times2$ $\mathbf{q}$-point grid of the 16 atom supercell, we interpolate the difference between the $\Gamma$ point SSCHA auxiliary dynamical matrix and the harmonic result at this point to the finer $2\times2\times2$ grid. Adding the harmonic result back to the interpolated result, we have obtained the anharmonic phonons in the $2\times2\times2$ grid. The anharmonic $\alpha^2F(\omega)$ is obtained by combining the calculated electron-phonon matrix elements, with the eigenvalues and eigenfrequencies of the dynamical matrix defined by the SSCHA auxiliary force constants. Considering the lack of symmetry in many of the studied structures, a large number of SSCHA configurations are needed to converge the calculation of the Hessian of the SSCHA free energy \cite{bianco2017second}, and, thus, we use the eigenvalues of the dynamical matrix defined by the auxiliary force constants $\boldsymbol{\Phi}$ to build the anharmonic $\alpha^2F(\omega)$ and not the eigenvalues defined by the Hessian of the free energy. The anharmonic $T_c$ calculated with one or other phonons is not expected to be very different, as it happens in LaH$_{10}$ \cite{errea2020quantum}. The superconducting critical temperature was obtained by solving the Allen-Dynes modified equation \cite{allen1975transition} with different values of the so-called Coulomb pseudopotential $\mu^*$.

\section{Results}

\label{sec:results}

We start our analysis of possible metastable states with partial tetrahedral sites occupation by creating  configurations in the 2$\times$2$\times$2 supercell by randomly occupying tetrahedral and octahedral interstitial sites in the fcc Pd lattice as described above, assuming a lattice parameter of 7.814 $a_0$, the equilibrium lattice parameter obtained in the Born-Oppenheimer classical energy surface for PdH with full octahedral occupation within PBE and without considering the zero point motion. For each initial number of atoms in octahedral sites, we create 10 random structures. As shown in Fig. \ref{energies}, these structure have a wide range of Born-Oppenheimer energies, with differences up to almost 0.4 eV per PdH. This shows that arranging H atoms at different interstitial sites can considerably change the total energy of the system, showing that H-H and Pd-H interactions are strongly dependent on the site of hydrogen atoms. For this lattice parameter (7.814 $a_0$), the structure with lowest Born-Oppenheimer energy is the one with full octahedral occupation ($n=8$). It is interesting to remark that, as illustrated in the bottom panel of Fig. \ref{energies}, these random structures have a very different pressure, which increases with the number of H atoms occupying tetrahedral sites, reaching a pressure of 160 kbar for the fully tetrahedral structure ($n=0$). Thus, structures with tetrahedral sites  occupied have a larger lattice parameter.        

As mentioned above, once tetrahedral interstitial sites start being occupied, the symmetry of the crystal is reduced and atoms are not fixed to their sites by symmetry anymore. We first relax these structures to the Born-Oppenheimer energy minima only modifying the internal positions, without modifying the fcc primitive lattice vectors and keeping the 7.814 $a_0$ lattice parameter. These reduces considerably the energies of the structures with tetrahedral sites occupied ($n<8$) with respect to the full octahedral occupation, but still the full octahedral site structure remains the lowest energy one. However, considering that the $n<8$ structures still are subject to a positive pressure with this unit cell, the energy is not the appropriate thermodynamic quantity for the comparison. We thus fully relax these structures to the Born-Oppenheimer minima with a target pressure of 0 kbar, with a tolerance of about 10 kbar. Most of the cell relaxations are performed by only adjusting the length of the lattice parameters, keeping fcc lattice vectors. In order to study the possibility of symmetry breaking of the cell, we also relaxed some structures allowing full adjustment of the lattice vectors with no restrictions. After the cell relaxations the pressure of all structures is comparable and the energy becomes the right thermodynamic variable. Interestingly, structures with partial or even full tetrahedral occupation have a lower energy than the full octahedral structure, remarking that tetrahedral and octahedral occupation of interstitial sites are very competitive in energy, as previously described in the literature \cite{elsasser1991vibrational,caputo2003h}.   

\begin{figure*}
\begin{center}
\addtolength{\tabcolsep}{5pt}
\begin{tabular}{c|c|c|c|c}
Structure&No relaxation&Internal relaxation&Cell relaxation&SSCHA relaxation\\\hline
\#1&\includegraphics[scale=0.2,trim={0 -10 0 -10}]{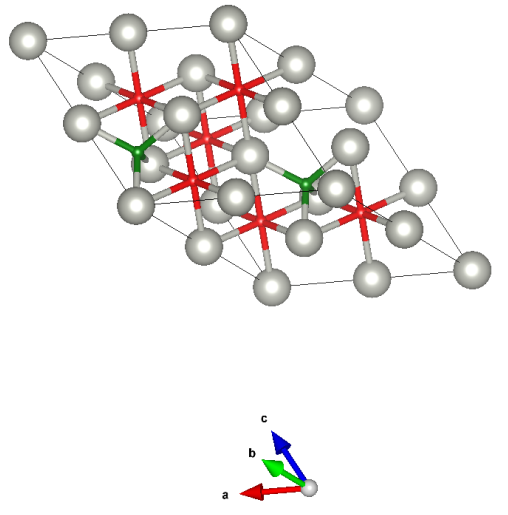}&\includegraphics[scale=0.2]{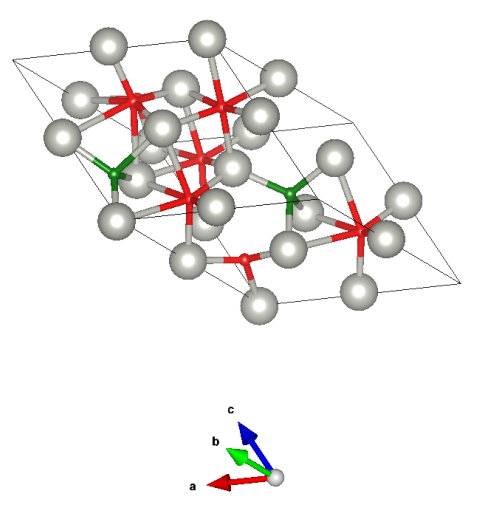}&\includegraphics[scale=0.2]{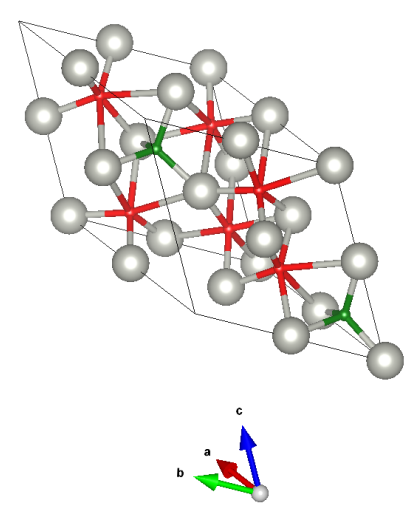}&\includegraphics[scale=0.2]{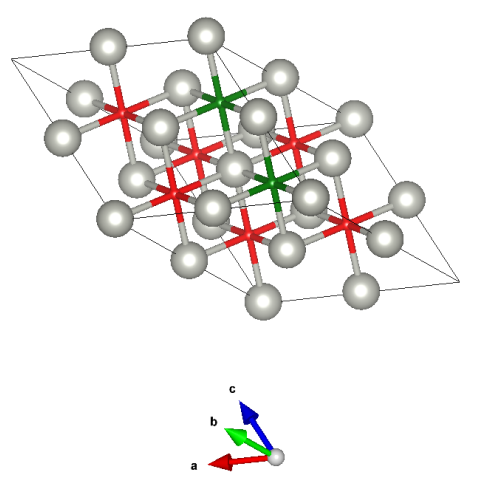}\\\hline
\#2&\includegraphics[scale=0.2,trim={0 -10 0 -10}]{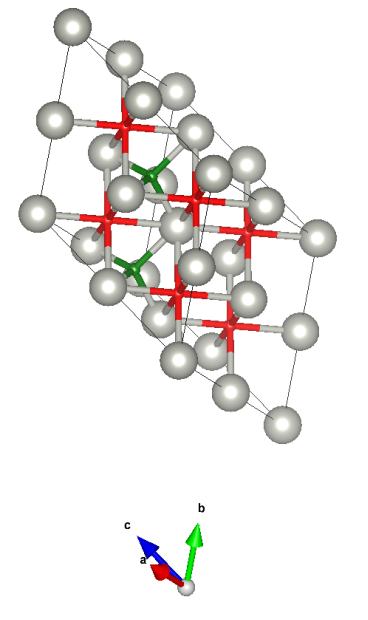}&\includegraphics[scale=0.2]{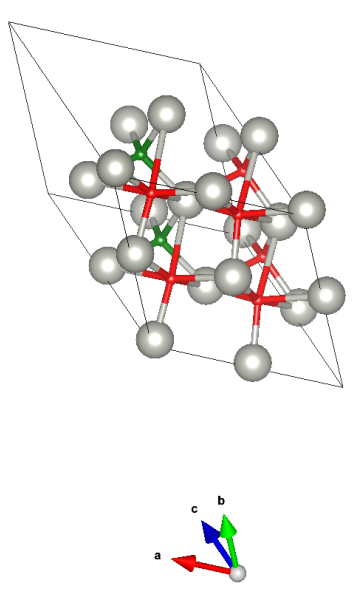}&\includegraphics[scale=0.2]{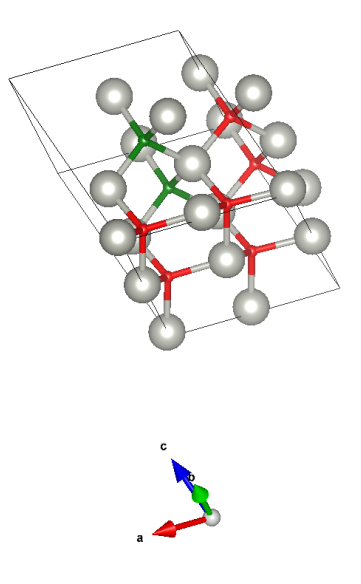}&\includegraphics[scale=0.2]{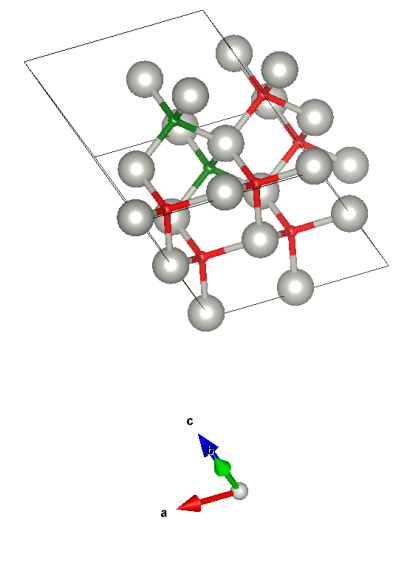}\\\hline
\#3&\includegraphics[scale=0.2,trim={0 -10 0 -10}]{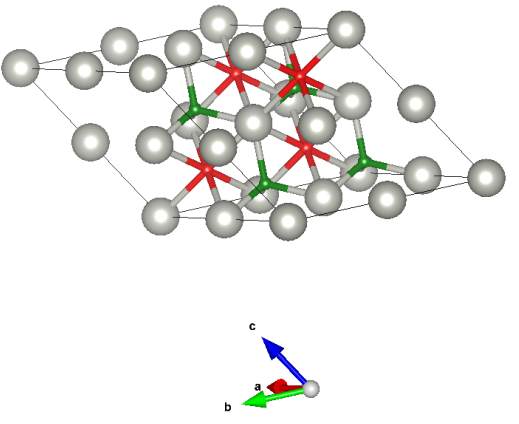}&\includegraphics[scale=0.2]{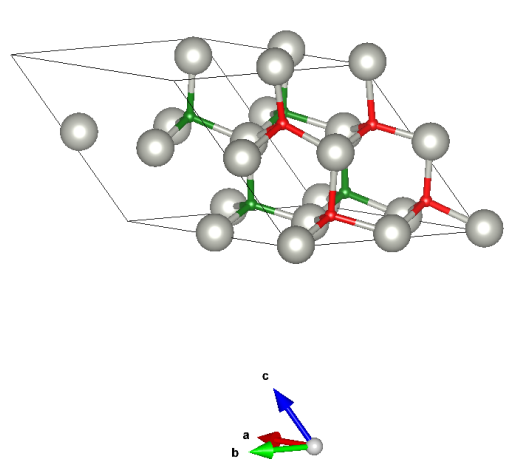}&\includegraphics[scale=0.2,trim=0 -15 0 -15]{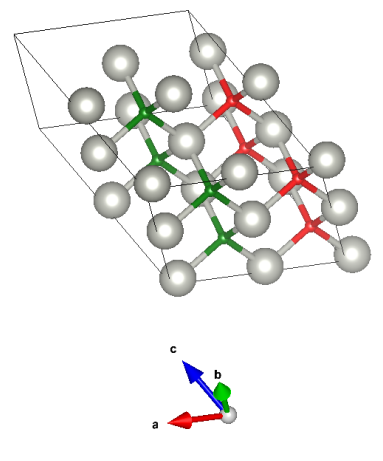}&\includegraphics[scale=0.2,trim={0 -10 0 -10}]{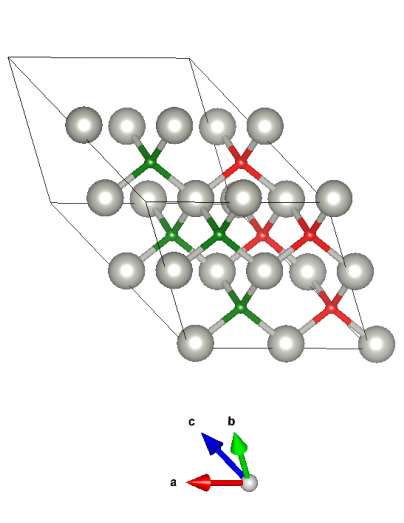}\\\hline
\#4&\includegraphics[scale=0.2,trim={0 -10 0 -10}]{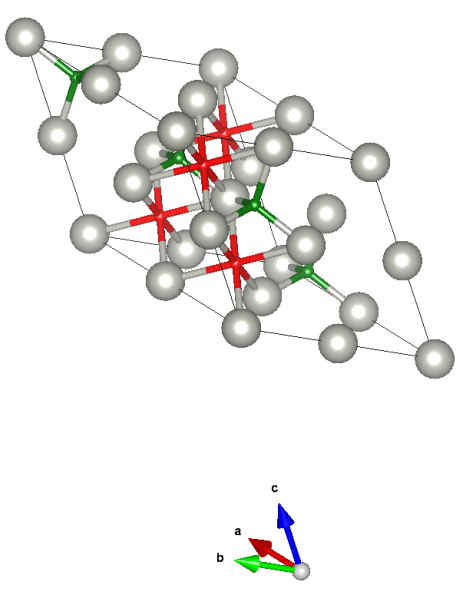}&\includegraphics[scale=0.2]{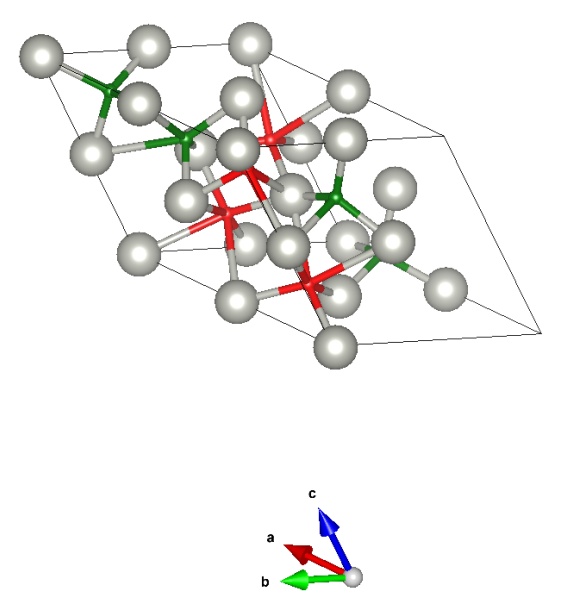}&\includegraphics[scale=0.2,trim={0 -10 0 -10}]{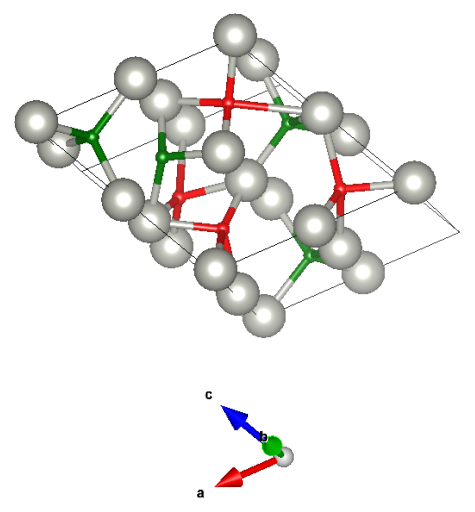}&\includegraphics[scale=0.2]{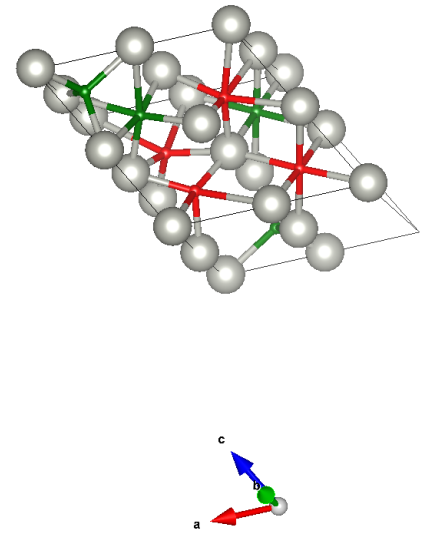}\\\hline
\#5&\includegraphics[scale=0.2]{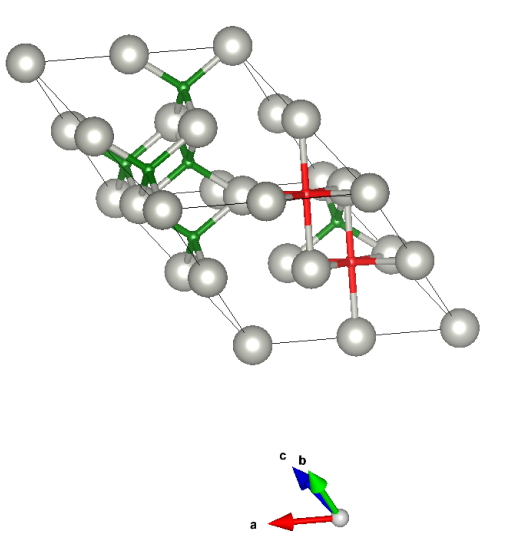}&\includegraphics[scale=0.2]{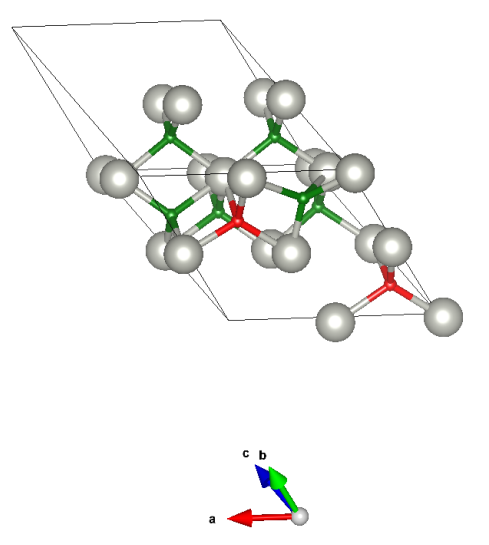}&\includegraphics[scale=0.2,trim={0 -10 0 -10}]{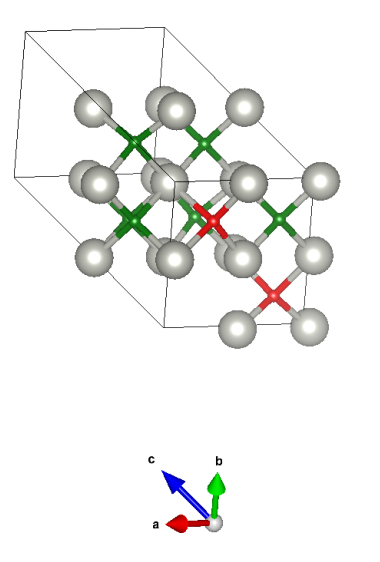}&\includegraphics[scale=0.2]{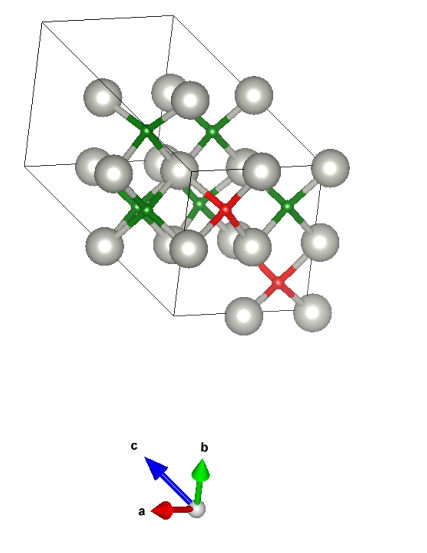}
\end{tabular}
	\caption{Graphical visualization of the structures, ordered from structure \#1 to \#5. For every structure, we present from left to right the non-relaxed starting structure (No relaxation), the relaxed structure  without modifications of the cell (Internal relaxation), the relaxed structure obtained by also relaxing the unit cell (Cell relaxation), and the structure obtained with the quantum relaxation given by the SSCHA (SSCHA relaxation). We mark in red the hydrogen atoms in octahedral sites at the beginning of the calculation and in green the ones in tetrahedral sites. The atoms keep their initial color for the other relaxations. The bonds plotted have a cutoff distance of $\sim 4.7 a_0$.}
	\label{structures}
\end{center}
\end{figure*}

The evolution of the classical relaxation of some of these structures is illustrated in Fig. \ref{structures}. The structures change strongly from their initial starting point, which addresses the shallowness of the energy barriers between different interstitial sites \cite{elsasser1991vibrational,blanco-rey2014electonic}. Some of the calculations show that, despite departing with some H atoms at octahedral sites, the final structures only contain H atoms at tetrahedral sites, like those labeled as \#2, \#3, and \#5. 
Structures \#1, \#3, \#4, and \#5 were relaxed keeping the fcc cell vectors and the lattice parameters increased between a 1.5\% and 2.5\% depending on the case. The structural modifications mainly involve hydrogen atoms in these cases, as Pd atoms remain with an arrangement that is not far from the perfect fcc. Considering that H atoms are invisible to x-ray diffraction experiments, distinguishing these structures from the full octahedral configuration would be difficult by x-ray, only an expansion of the lattice should be detected for those cases with occupied tetrahedral sites. For structure \#2 the situation is different, as the unit cell vectors are allowed to fully change: at the end of the classical cell relaxation a monoclinic lattice is obtained, with  two lattice vectors that dilate by factors of 1.4\% and 2.4\%, respectively, and the other vector that contracts by a factor of 0.5\%. The angles shift to $59^\circ$, $74^\circ$ and $60^\circ$ between relaxed unit cell vectors. Due to the lattice change, the structure should be distinguishable from the $Fm\bar{3}m$ structure by x-ray diffraction.

Considering the large anharmonicity at play in PdH, at least in the fully octahedral configuration \cite{errea2013first}, we further relax some of the structures obtained after the  cell Born-Oppenheimer classical relaxation considering ionic quantum effects and anharmonicity within the SSCHA. Due to the large computational effort to structurally relax in the quantum energy landscape these structures that lack symmetry, we limit ourselves to relax the five  representative structures labeled from \#1 to \#5 in all figures. Despite ionic quantum effects and anharmonicity can impact the cell parameters \cite{errea2020quantum,monacelli2018pressure,hou2021quantum,hou2021strong}, considering the additional large computational effort required to relax the cell parameters for these systems with no symmetry, here we do not relax the unit cell further and just relax the internal centroid positions $\bRcal$ in the quantum energy landscape. However, the stress tensor calculated including ionic quantum effects \cite{monacelli2018pressure} in these structures shows that there is a weak anisotropy in the stress tensor for any of the five structures, irrespective of whether the cell remains fcc-like or not. This suggests that fcc lattices can be realized in this system even if the symmetry is broken by the interstitial sites occupied by hydrogen, supporting the treatment given to the cell relaxation.

The impact of ionic quantum effects on the centroid positions is structure dependent. While some structures are radically changed by quantum effects, others do not exhibit drastic changes (see Fig. \ref{structures}). 
Structure \#1, which starts with 6 H atoms in octahedral sites and 2 in tetrahedral sites, deforms considerably in the classical relaxation with fixed cell and relaxing the cell, but still keeps hydrogens at the same interstitial sites. This is changed by the quantum SSCHA relaxation, which shifts the atoms in tetrahedral sites to octahedral sites, recovering the full octahedral configuration. Structure \#2, which also starts with 6 octahedral and 2 tetrahedral sites, keeps those occupations after the classical internal relaxation, but the cell relaxation pushes all atoms into tetrahedral sites. The posterior quantum SSCHA relaxation barely affects the structure, keeping all the H atoms at tetrahedral sites. Structure \#3, starting with four octahedral and four tetrahedral sites, already with the first internal classical relaxation transforms into a full tetrahedral occupation, which does not evolve further after the classical cell  and the quantum SSCHA relaxations. Structure \#4, which start as well with an equal occupation of octahedral and tetrahedral sites, is particularly interesting as each relaxation induces a large change in the position of hydrogen atoms. Remarkably, ionic quantum effects stabilize a structure with mixed occupation of tetrahedral and octahedral sites. Finally, structure \#5, which start with the occupation of 6 tetrahedral and 2 octahedral sites, is similar to structure \#2 since already at the internal relaxation all H atoms shift to a tetrahedral site and remain unaltered by the subsequent relaxations. All these calculations show that the quantum energy landscape in the PdH system is complex and many possible metastable states are possible also when the zero point energy and anharmonicity are considered in the calculations. Regardless of the initial occupation of interstitial sites, our calculations show that stable or metastable states with full tetrahedral or octahedral occupations, as well as with mixed occupation of these sites, are possible.       

\begin{figure}
	\centering
	\includegraphics[width=\columnwidth]{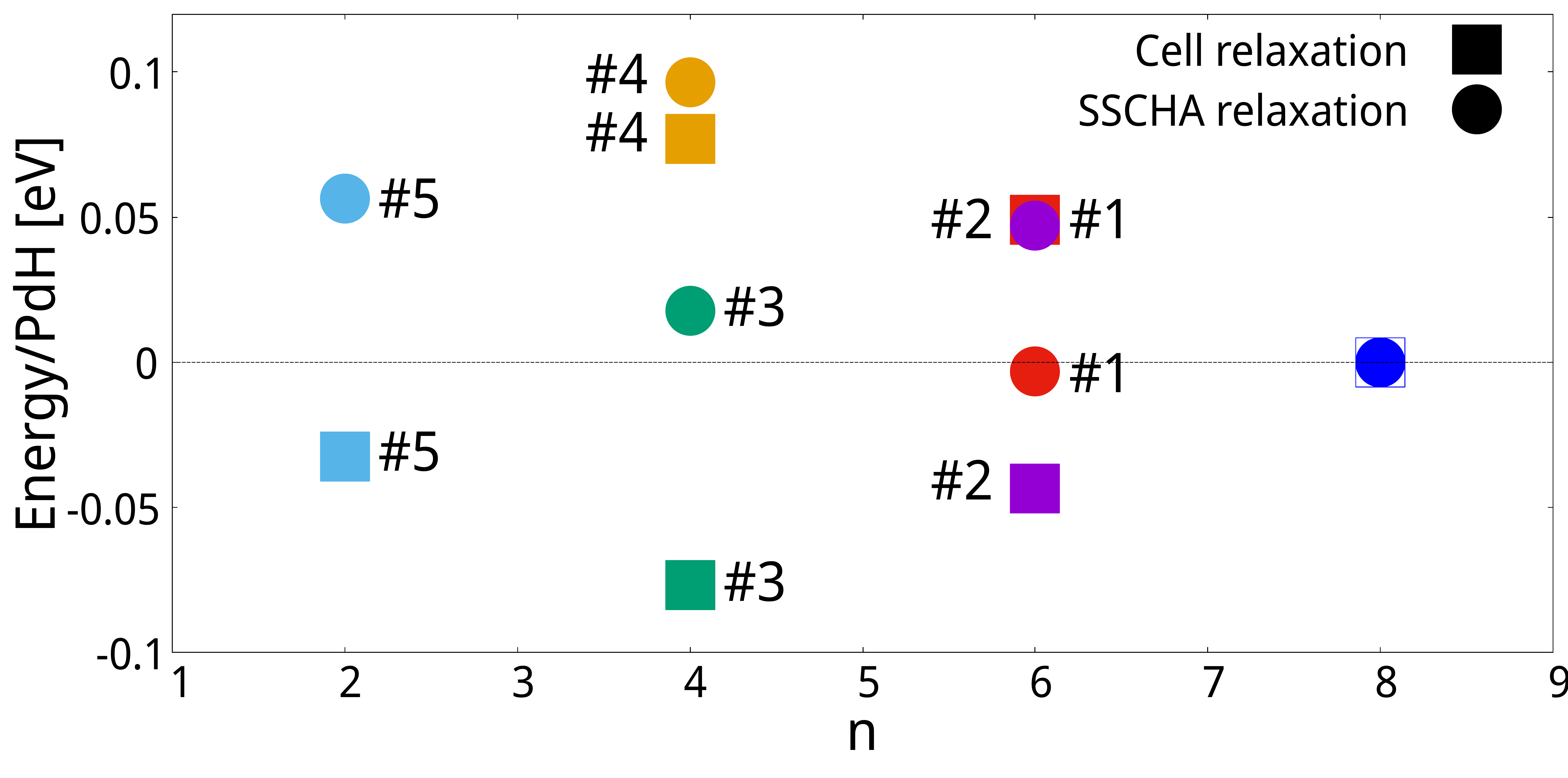}
	\caption{Comparison between the energies obtained  with the classical relaxation including the unit cell (Cell relaxation) and with the quantum relaxation within the SSCHA (SSCHA relaxation) for the five selected structures. The zero of the energy is taken to be the energy of the structure with 8 occupied octahedral sites. The number on the right of each data point indicates the structure. The structure are characterized by $n$, the initial number of occupied octahedral sites.}
	\label{graphzpe}
\end{figure}

Our SSCHA calculations can also analyze which of these states is the ground state structure once the ionic zero point energy and anharmonicity is considered in the calculations. As shown in Fig. \ref{graphzpe}, despite structures \#2, \#3, and \#5
have a lower Born-Oppenheimer energy than the high-symmetry full octahedral structure, after the SSCHA relaxations the energy including ionic quantum anharmonic effects of these structures becomes larger. Consistently, structure \#1, which evolves into the full octahedral occupation, reaches the same energy as the high symmetry $n=8$ case. Thus, our calculations suggest that the structure with full occupation of octahedral sites is the ground state, as reported experimentally \cite{WORSHAM1957neutron}, thanks to lattice quantum effects and anharmonicity. However, metastable states with full or partial occupation of tetrahedral sites exist within 0.1 eV per PdH.

\begin{figure*}
    \noindent 
	\begin{minipage}{0.45\linewidth}
	\includegraphics[width=1.0\linewidth]{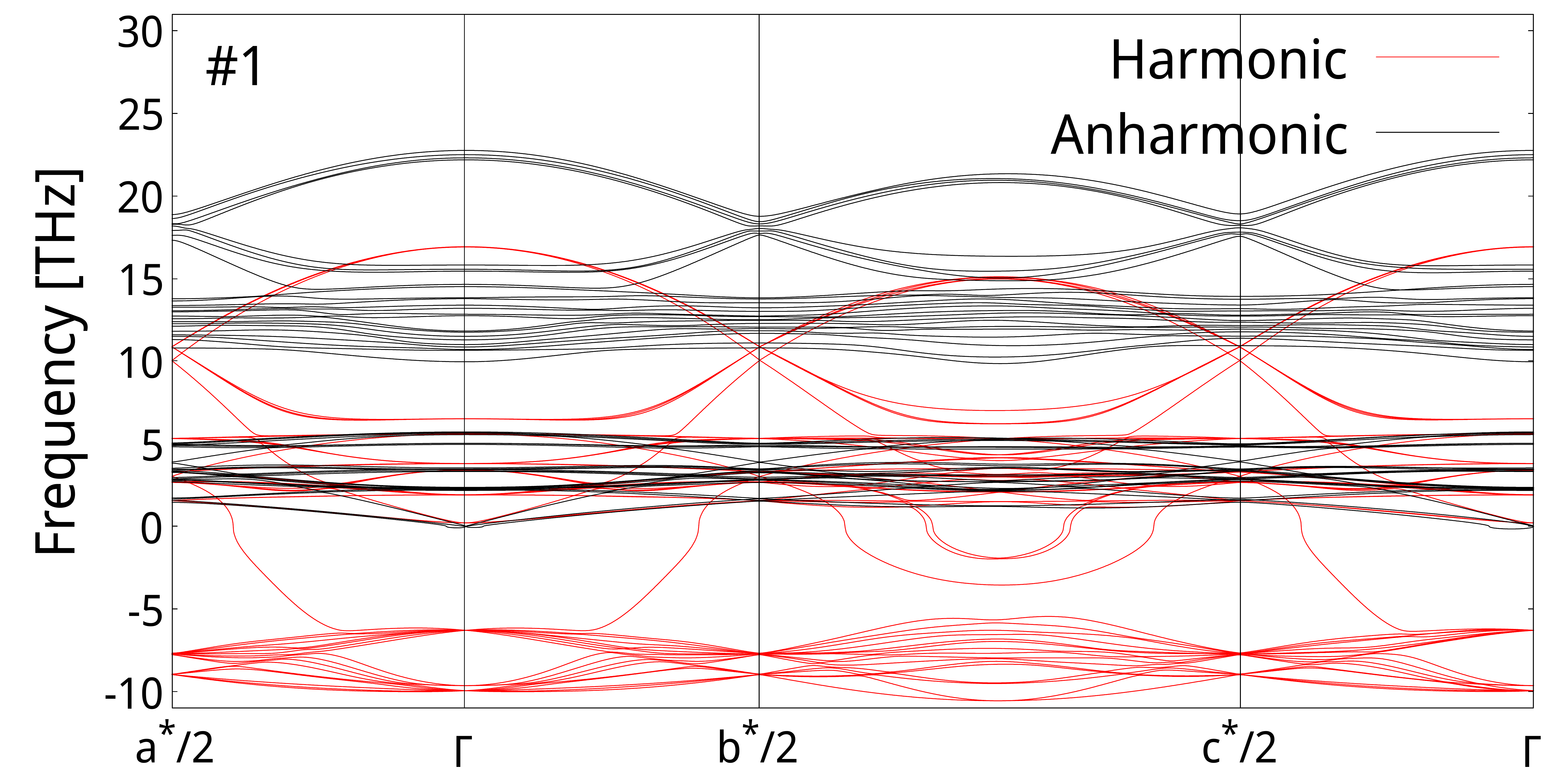}
	\includegraphics[width=1.0\linewidth]{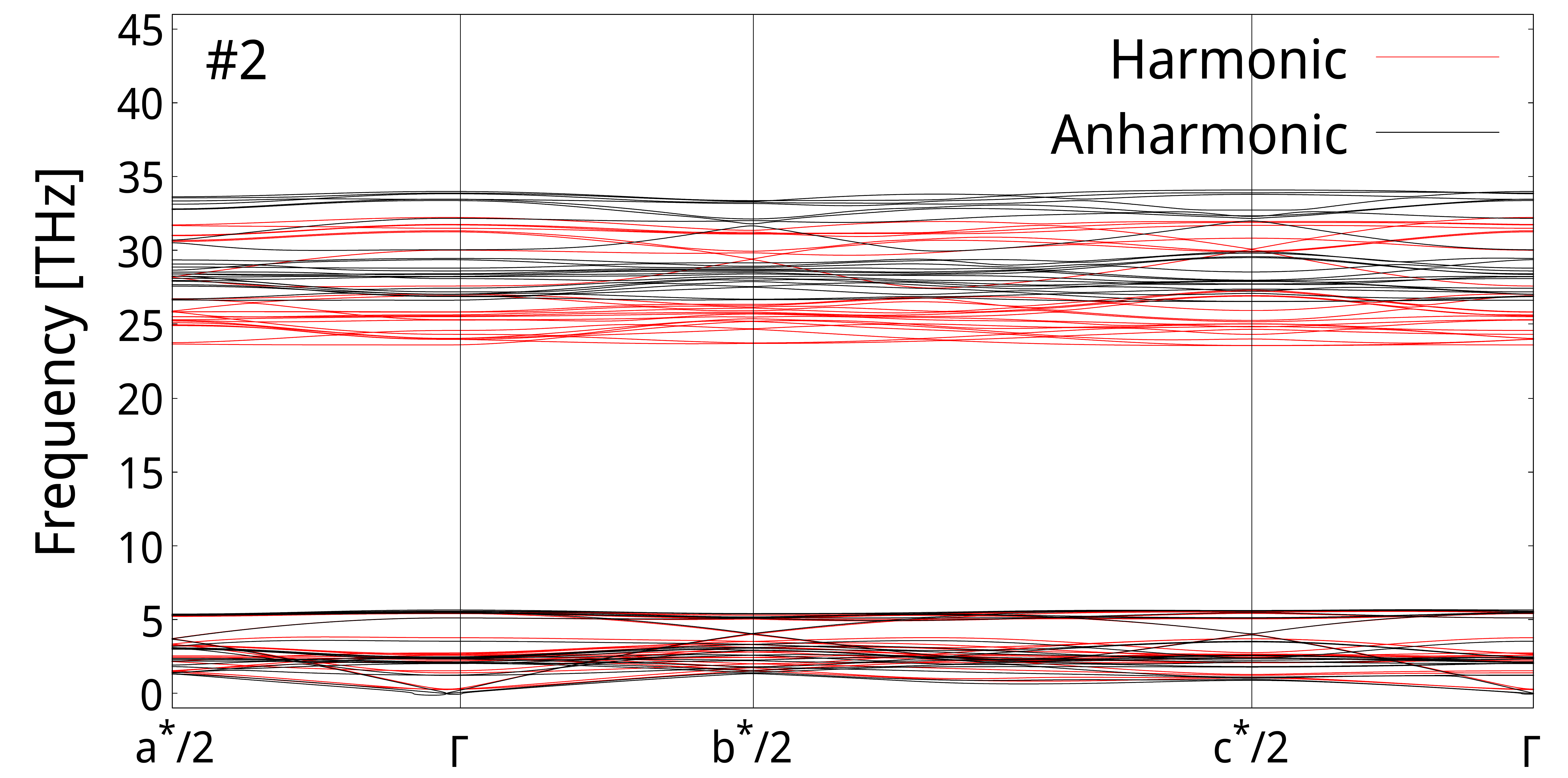}
	\includegraphics[width=1.0\linewidth]{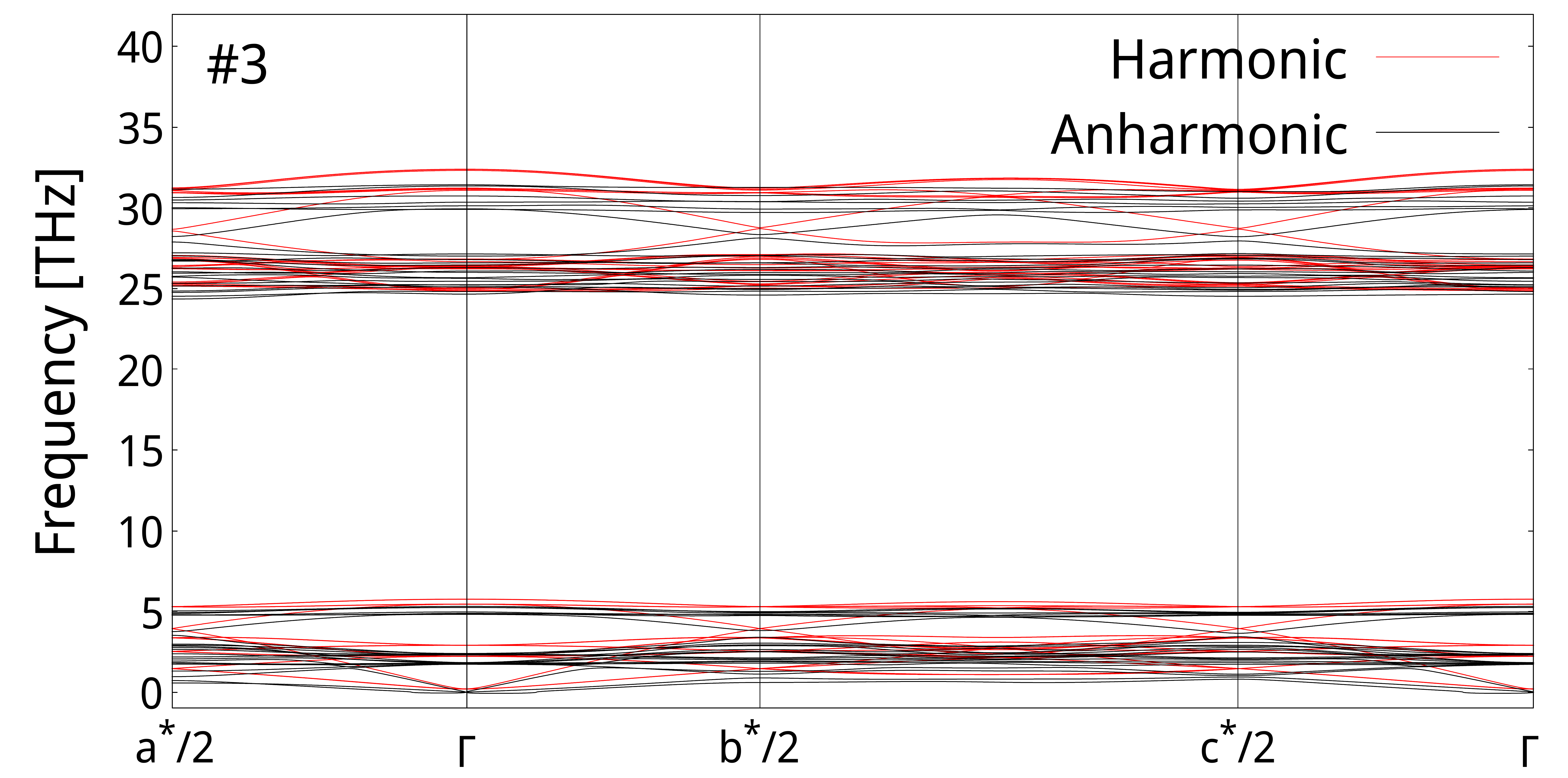}
	\includegraphics[width=1.0\linewidth]{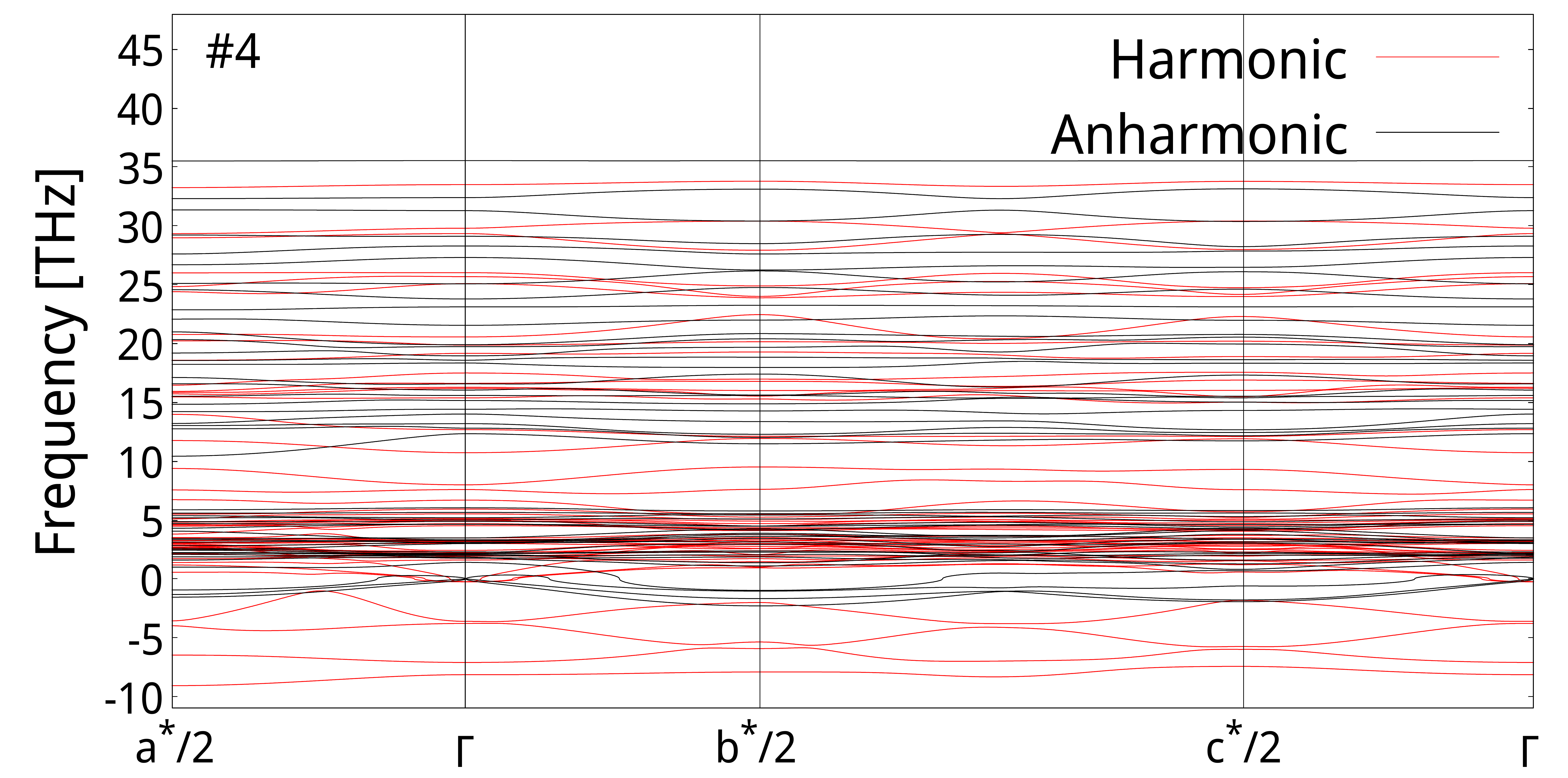}
	\includegraphics[width=1.0\linewidth]{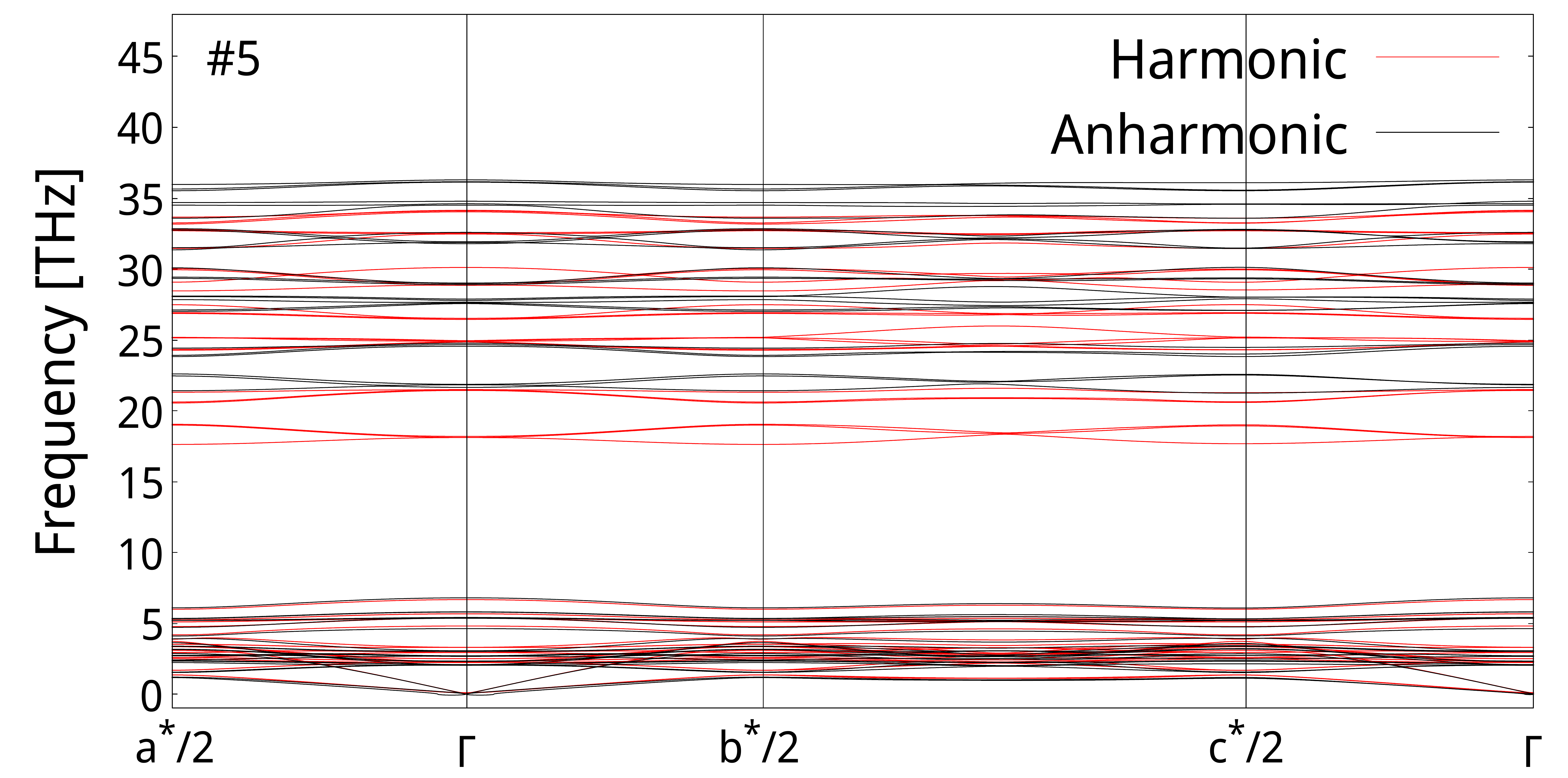}
	\caption{Phonon spectra obtained in the harmonic and anharmonic calculations for the five selected structures in the structure obtained with the SSCHA relaxation. The vectors $a^*$, $b^*$, and $c^*$ are the basis vectors of the reciprocal lattice.}
	\label{phononbands}
\end{minipage}%
\hfill%
    \noindent 
	\begin{minipage}{0.45\linewidth}
	\includegraphics[width=1.0\linewidth]{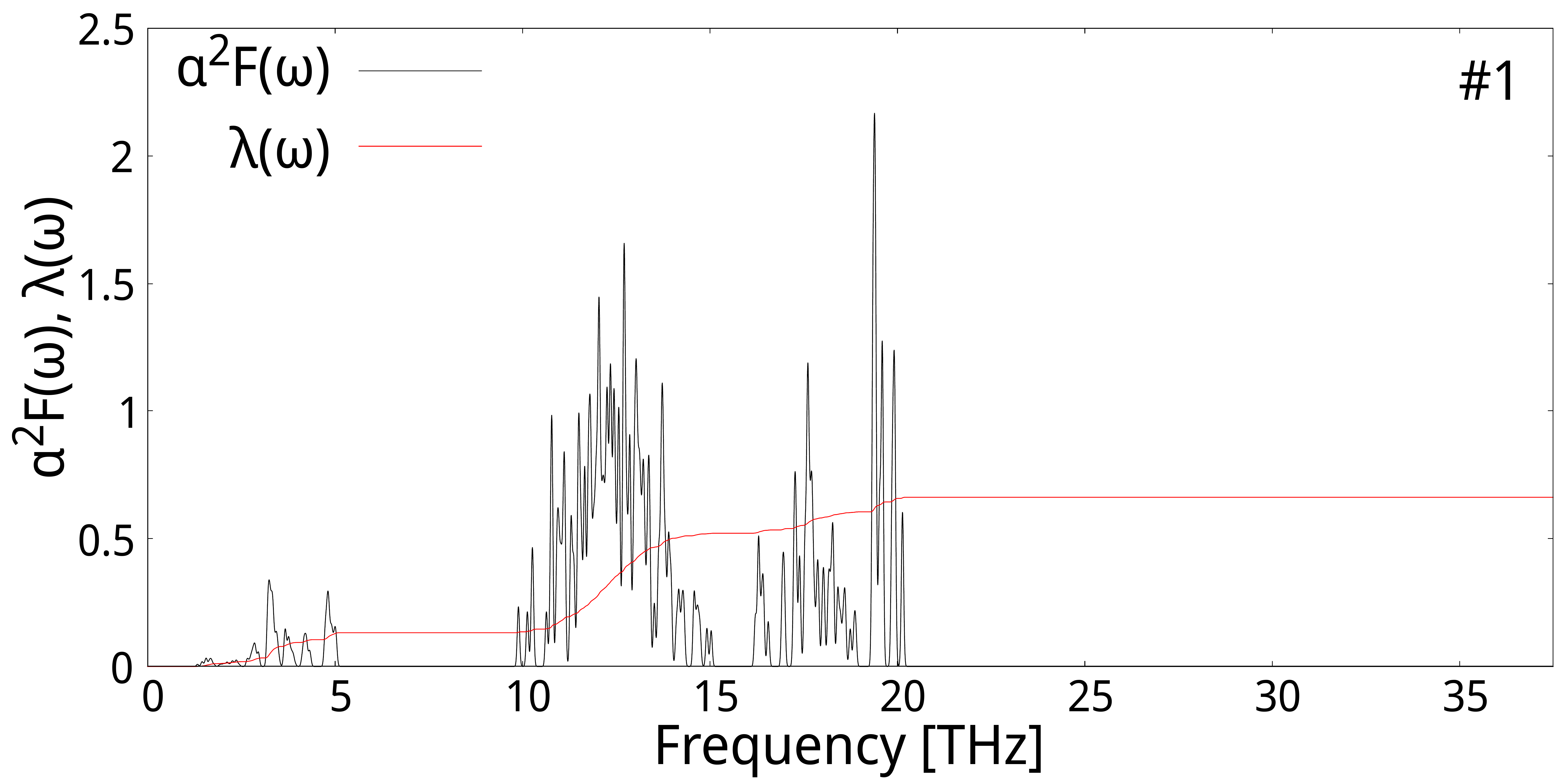}
	\includegraphics[width=1.0\linewidth]{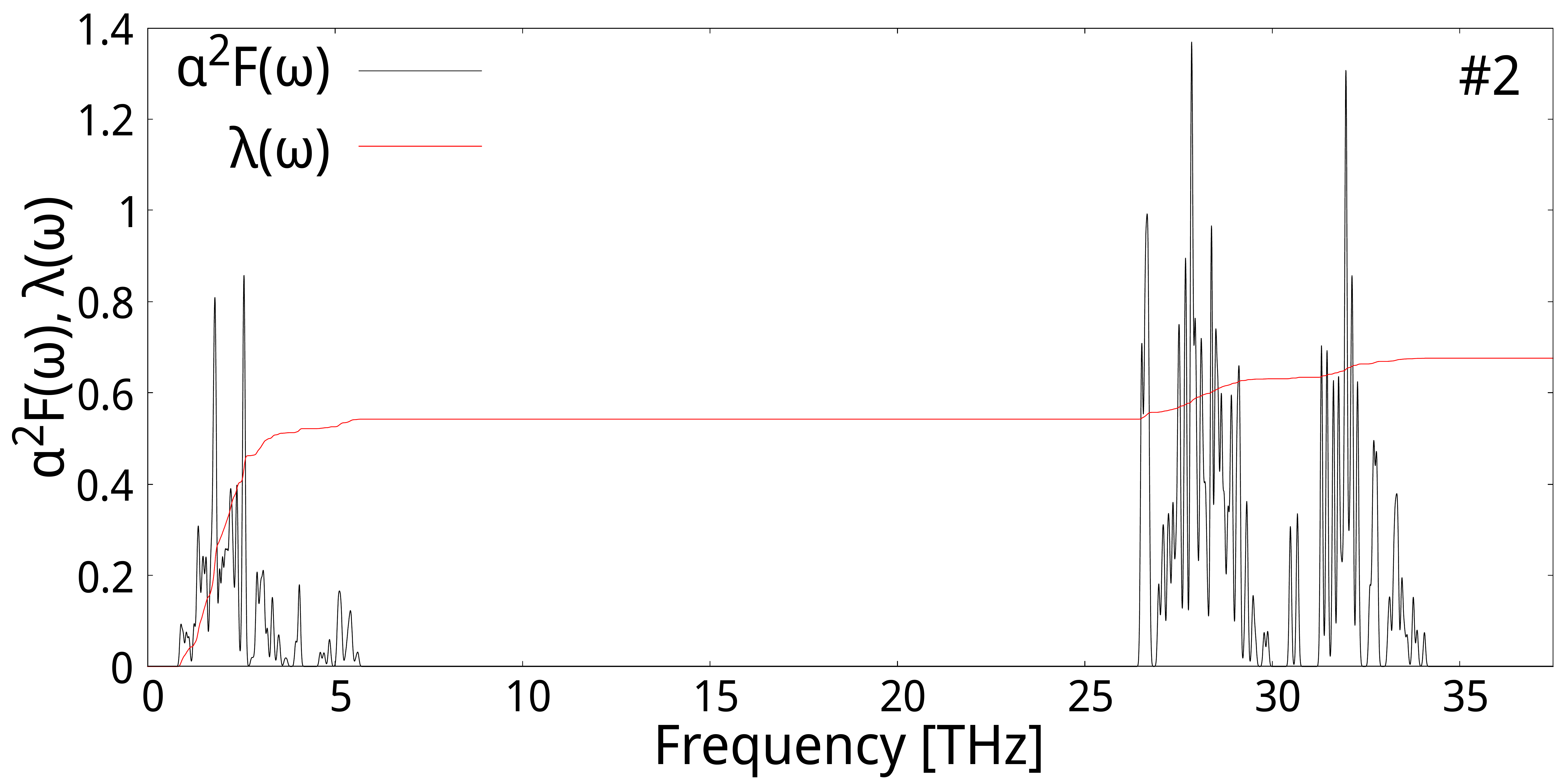}
	\includegraphics[width=1.0\linewidth]{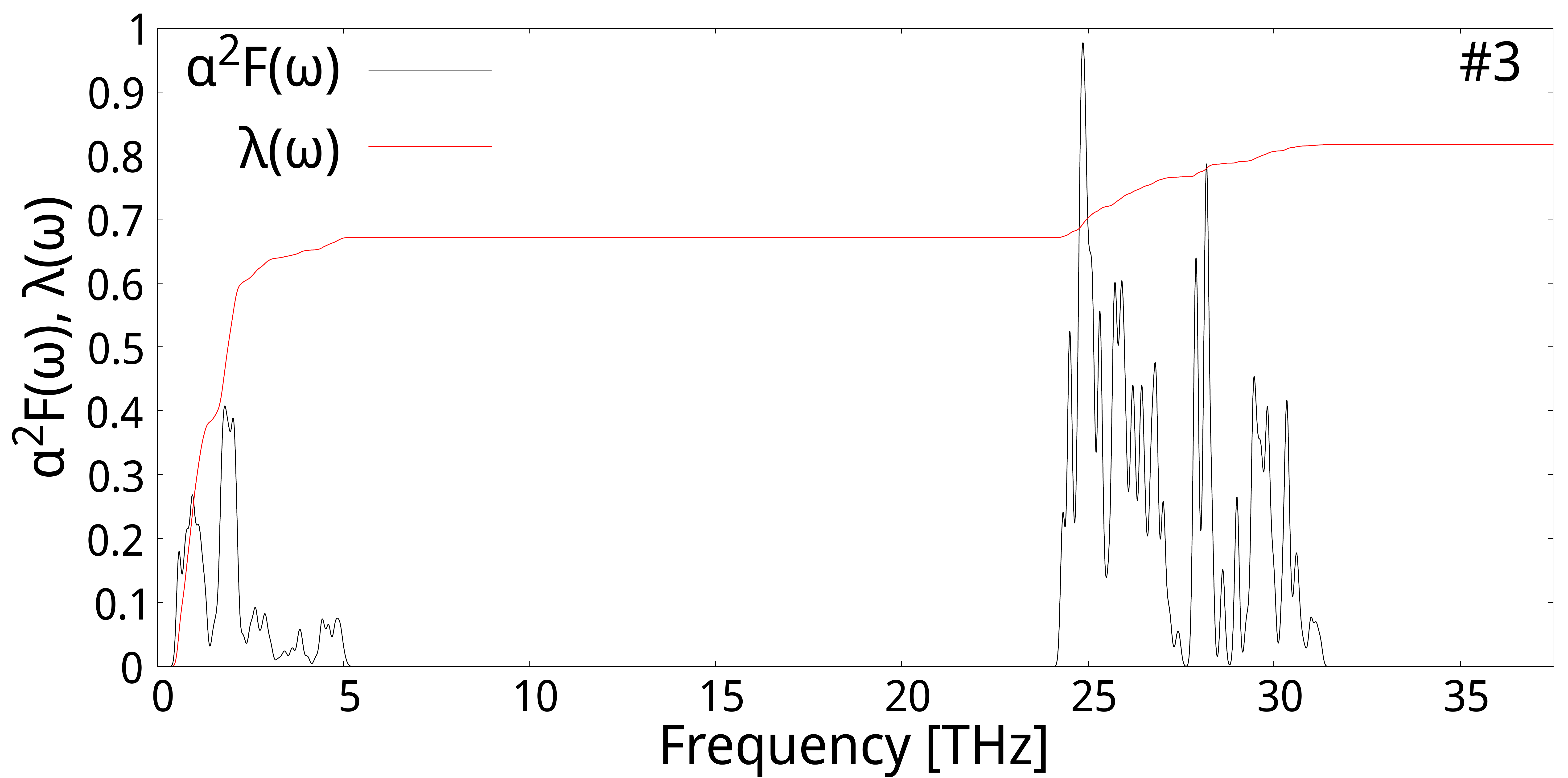}
	\includegraphics[width=1.0\linewidth]{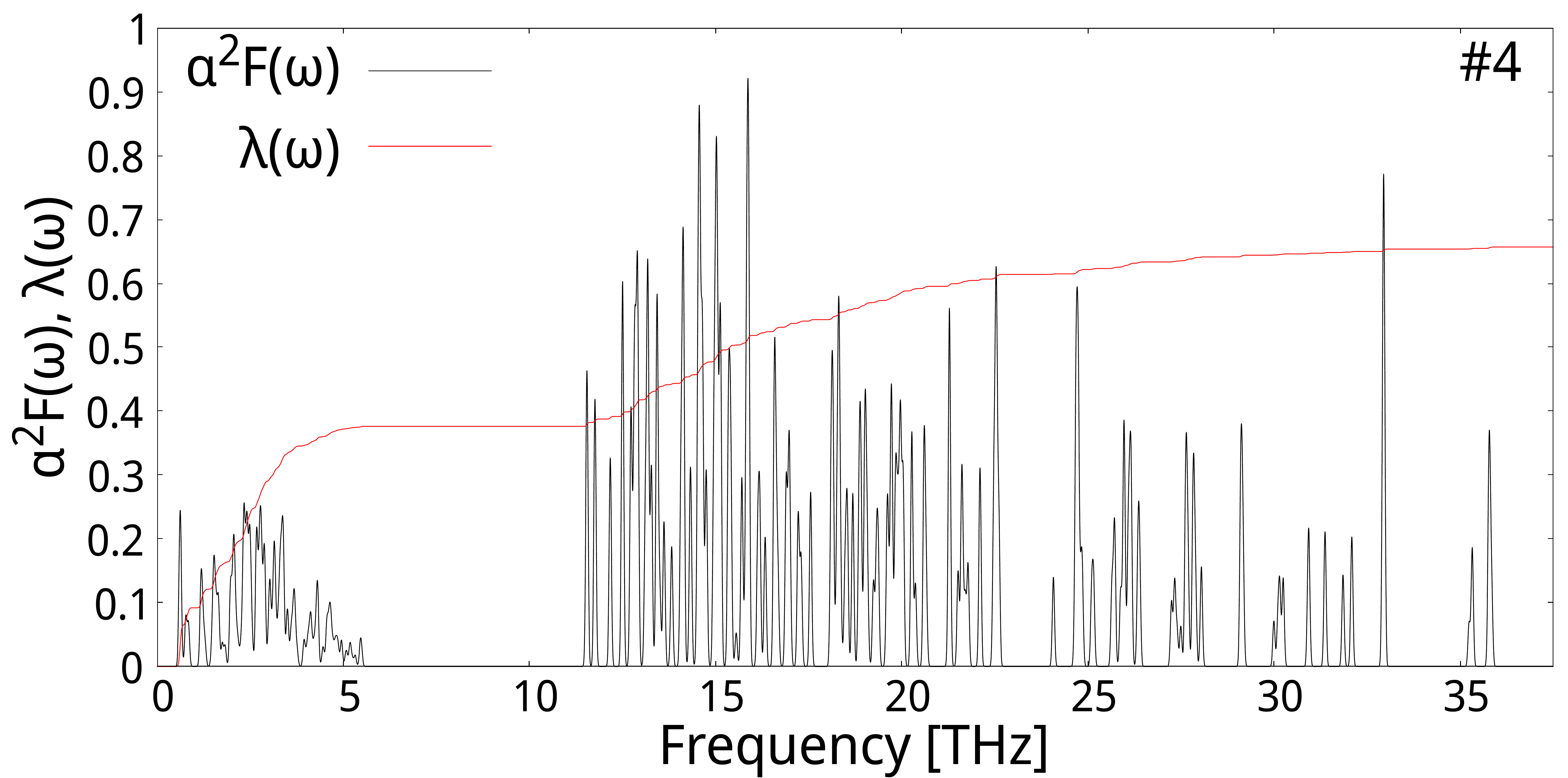}
	\includegraphics[width=1.0\linewidth]{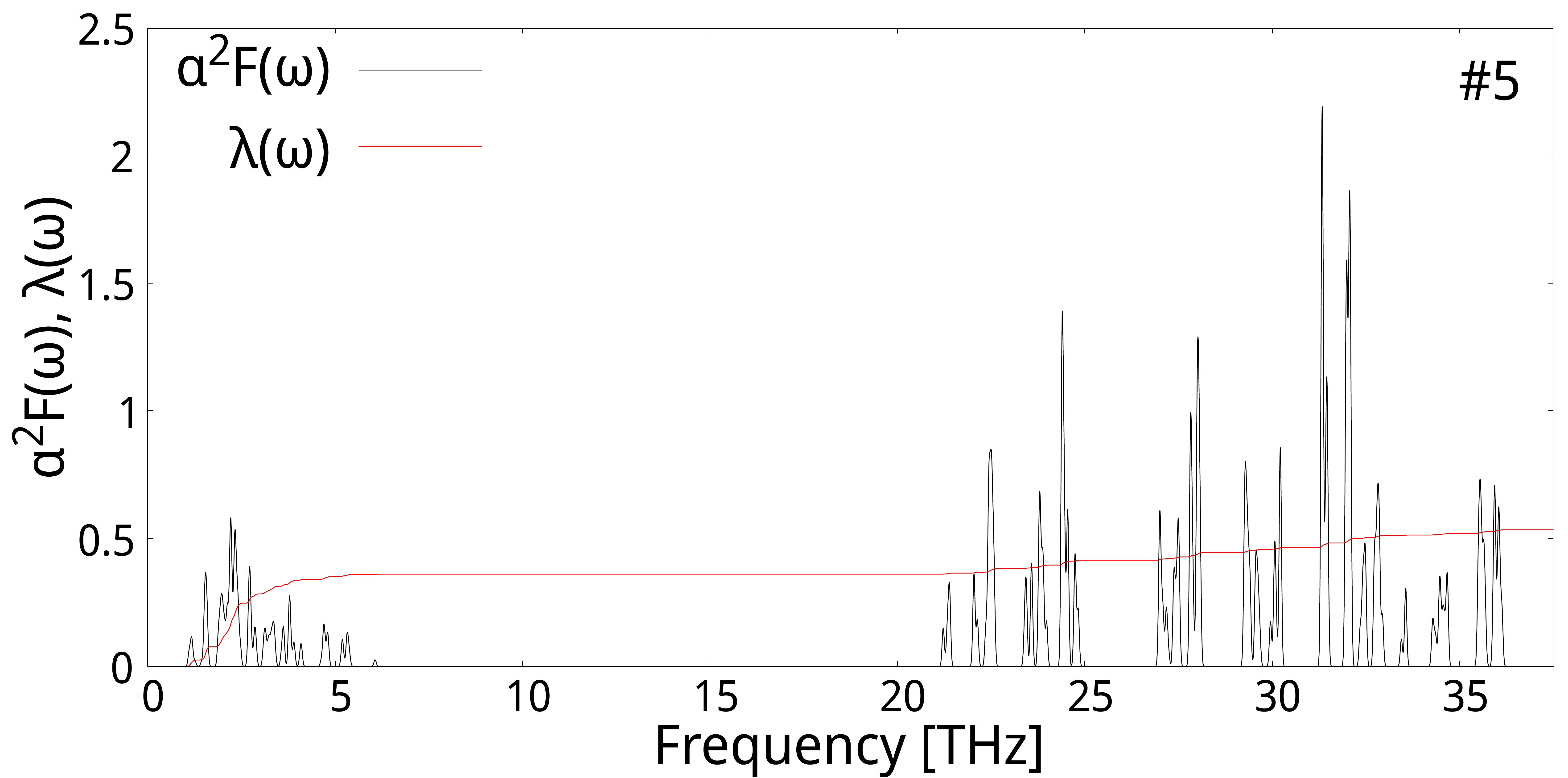}
	\caption{Eliashberg function $\alpha^2F(\omega)$ and integrated electron-phonon coupling constants $\lambda(\omega)$ for the five selected structures in the structure obtained with the SSCHA relaxation. Anharmonic phonons are used in this calculation.}
	\label{a2f}
\end{minipage}
\end{figure*}

\begin{table*}
	\centering
	\begin{tabular}{c|c|c|c|c|c|c}
Structure&\multicolumn{2}{c|}{Harmonic $T_c$ (K)}&\multicolumn{2}{c|}{SSCHA $T_c$ (K)}&Initial  octahedral&Final octahedral \\
&$\mu^*=0.085$&$\mu^*=0.130$&$\mu^*=0.085$&$\mu^*=0.130$&&\\\hline
\#1&$\times$&$\times$&17.7&11.5&6&8\\\hline
\#2&5.5&3.2&6.1&4.0&6&0\\\hline
\#3&2.2&0.9&5.9&4.3&4&0\\\hline
\#4&$\times$&$\times$&12.2&8.7&4&Mixed\\\hline
\#5&1.5&0.4&4.9&2.6&2&0
	\end{tabular}
	\caption{Superconducting critical temperatures calculated with different number of $\mu^*$ for the selected structures after the quantum SSCHA relaxation, computed with harmonic dynamical matrices and SSCHA auxiliary dynamical matrices. We also note the number of octahedral occupied positions before any relaxation and after the final SSCHA relaxation. Structures \#1 and \#4 do not have a harmonic value due to the presence of imaginary phonon frequencies at this level of theory.}
	\label{critical}
\end{table*}

The reason why quantum effects affect one structure more than others is reflected in the phonon spectra shown in Fig. \ref{phononbands}. For the final structure obtained after the SSCHA relaxation we calculate the phonon spectra in the harmonic approximation and from the SSCHA auxiliary dynamical matrices. Structures \#2, \#3, and \#5 are barely affected by anharmonicity, while structures \#1 and \#4 are strongly anharmonic. This is consistent with the fact that the latter structures are the only ones that are considerably modified by quantum anharmonic effects. Interestingly, structure \#1 has large imaginary phonon modes in the harmonic approximation, as it has previously been described for the high-symmetry $Fm\bar{3}m$ structure \cite{errea2013first}. The same occurs to structure \#4, which also has H atoms in octahedral sites. The analysis of the phonon spectra clearly indicates thus that anharmonicity is related to octahedral sites and not so much to tetrahedral sites, for which the potential seems rather harmonic. Hydrogen vibrations in tetrahedral sites are much harder, with energies above approximately 25 THz, while hydrogen atoms in octahedral sites vibrate with energies between approximately 10 and 20 THz. This is the reason why the structure with mixed octahedral and tetrahedral occupation (\#4) has the broadest range of phonon frequencies. The differences between the structures and the lack of degeneracies in the phonon spectra are a consequence that all structures are not identical as no symmetries have been imposed.   

The existence of metastable states with full and partial occupation of tetrahedral sites suggests that it may be possible to synthesize these structures with fast-cooling techniques, as suggested by Syed et al. \cite{syed2016superconductivity}. It has to be seen, however, whether these structures do increase the superconducting critical temperature with respect to the stable full octahedral configuration, as hypothesized in Ref. \cite{syed2016superconductivity}. We calculate here the electron-phonon interaction and the $T_c$ for the five structures analyzed previously, with the aim to address this  important point. Fig. \ref{a2f} shows the Eliashberg function, $\alpha^2F(\omega)$, and the integrated electron-phonon coupling constant, $\lambda(\omega)$, for the anharmonic SSCHA calculation. From the figures we can deduce that the contribution of the high-energy H-character modes to the electron-phonon coupling constant is only sizable when octahedral sites are occupied, such as in structures \#1 and \#4. It is thus not surprising that the largest calculated $T_c$ among these structures is the one with full octahedral occupation (\#1), with $T_c$ around 11.5 K with $\mu^*=0.13$, followed by the one with mixed occupation (\#4), with $T_c$ around 8.7 K for the same value of the Coulomb pseudopotential (see Table \ref{critical}). With the same parameters, the structures with full tetrahedral occupation have a lower critical temperature, between 2.6 and 4.3 K. These show that the metastable occupation of tetrahedral sites is not beneficial for superconductivity and that the highest critical temperature is attained with the stable structure with full octahedral occupation.

The critical temperature calculated for the full octahedral configuration here is overestimated with respect to previous calculations also including anharmonic effects \cite{errea2013first}. We attribute these differences to the fact that the lattice vectors here have not been relaxed in the quantum energy landscape, so that the lattice parameters in the calculations are different; a different exchange-correlation functional is used here; and the calculation here is performed in a non-symmetrized structure, not exactly with the $Fm\bar{3}m$ symmetry, in a 2$\times$2$\times$2 supercell. The previous calculations, thus, are performed in a more controlled way and are thus more reliable with respect to the absolute value of the calculated $T_c$, but the result obtained here is solid: full or partial metastable occupation of tetrahedral sites reduces the critical temperature.

\section{Conclusion}
\label{sec:conclusions}

By analyzing many possible occupations of interstitial sites in palladium hydrides within DFT including ionic quantum and anharmonic effects, we conclude that, despite the configuration with all hydrogen atoms in octahedral sites is the ground state, partial and full tetrahedral sites can be occupied metastably. This opens the door to synthesizing structures with partial or full tetrahedral occupation, as suggested by recent experiments \cite{syed2016superconductivity,Pitt2003Tetrahedral,McLennan2008deuterium}. However, contrary to the hypothesis presented in Ref. \cite{syed2016superconductivity}, we demonstrate here that the occupation of tetrahedral sites does not increase the critical temperature of PdH, it suppresses it. Our results underline that quantum effects and anharmonicity are crucial in the thermodynamical (meta)stability of PdH structures, specially with respect to vibrations around the octahedral sites. The experiment measuring $T_c$ values above 50 K in metastable PdH phases \cite{syed2016superconductivity} is questioned by our theoretical calculations.

\section*{Acknowledgements}

A.M. acknowledges the support of Red Española de Supercomputación (RES) for CPU time (grant agreements FI-2020-1-0008 and FI-2021-2-0016) and PRACE for access to Joliot-Curie Rome at TGCC, France. We have received funding from the European Research Council (ERC) under the European Union’s Horizon 2020 research and innovation programme (grant agreement No 802533).

\bibliography{biblio}

\end{document}